\documentclass[times,twocolumn,final]{elsarticle}

\usepackage{medima}
\usepackage{framed,multirow}

\usepackage{amssymb}
\usepackage{latexsym}
\usepackage{amsmath}

\usepackage{url}
\usepackage{xcolor}
\usepackage{todonotes}
\usepackage{hyperref}
\usepackage{lineno}

\definecolor{newcolor}{rgb}{.8,.349,.1}

\journal{Medical Image Analysis}

\begin{document}
	
	\verso{Luyi Han \textit{et~al.}}
	
	\begin{frontmatter}
		
		\title{GAN-based disentanglement learning for chest X-ray rib suppression} 
		
		\author[1,2]{Luyi \snm{Han}}
		\author[3]{Yuanyuan \snm{Lyu}}
		
		\author[4]{Cheng \snm{Peng}}
		
		\author[5,6]{S.Kevin \snm{Zhou}\corref{cor1}}
		\cortext[cor1]{Corresponding author: S. Kevin Zhou;}
		\ead{s.kevin.zhou@gmail.com}
		
		\address[1]{Department of Radiology and Nuclear Medicine, Radboud University Medical Center, Geert Grooteplein 10, 6525 GA, The Netherlands}
		\address[2]{Department of Radiology, Netherlands Cancer Institute (NKI), Plesmanlaan 121, 1066CX, Amsterdam, The Netherlands}
		\address[3]{Z$^2$Sky Technologies Inc., Suzhou, 215123, China}
		\address[4]{Artificial Intelligence for Engineering and Medicine Lab, Johns Hopkins University, MD, USA}
		\address[5]{School of Biomedical Engineering \& Suzhou Institute for Advanced Research, University of Science and Technology of China, Suzhou, 215123, China}
		\address[6]{Key Lab of Intelligent Information Processing of Chinese Academy of Sciences (CAS), Institute of Computing Technology, CAS, Beijing, 100190, China}
		
		\received{-}
		\finalform{-}
		\accepted{-}
		\availableonline{-}
		\communicated{-}

		\begin{abstract}
			Clinical evidence has shown that rib-suppressed chest X-rays (CXRs) can improve the reliability of pulmonary disease diagnosis.
			However, previous approaches on generating rib-suppressed CXR face challenges in preserving details and eliminating rib residues.
			We hereby propose a GAN-based disentanglement learning framework called Rib Suppression GAN, or RSGAN, to perform rib suppression by utilizing the anatomical knowledge embedded in unpaired computed tomography (CT) images. In this approach, we employ a residual map to characterize the intensity difference between CXR and the corresponding rib-suppressed result.
			To predict the residual map in CXR domain, we disentangle the image into structure- and contrast-specific features and transfer the rib structural priors from digitally reconstructed radiographs (DRRs) computed by CT.
			Furthermore, we employ additional adaptive loss to suppress rib residue and preserve more details.
			We conduct extensive experiments based on 1,673 CT volumes, and four benchmarking CXR datasets, totaling over 120K images, to demonstrate that (i) our proposed RSGAN achieves superior image quality compared to the state-of-the-art rib suppression methods; (ii) combining CXR with our rib-suppressed result leads to better performance in lung disease classification and tuberculosis area detection.
		\end{abstract}
		
		
	\end{frontmatter}
	
	
	\section{Introduction}
	Chest X-ray (CXR), a 2D projection of a 3D scene from an X-ray source, and computed tomography (CT), which reconstructs a 3D scene based on a collection of 2D X-ray projections, are the widely used modalities to diagnose lung disease. 
	As compared to CT, CXR examinations induce up to 120 times lower radiation dose and are more affordable~\citep{CTdoes}.
	However, diagnosis with CXR alone is challenging and misses lung cancer findings~\citep{missedLungCancer}.
	That is because CXR represents a 2D projection of the 3D chest and contains overlapped anatomies and ambiguous structure details.
	Thereby, it is clinically significant to improve the diagnostic reliability of CXR for better clinical decision making.
	
	\begin{figure*}[t]
		\centering
		\includegraphics[scale=.9]{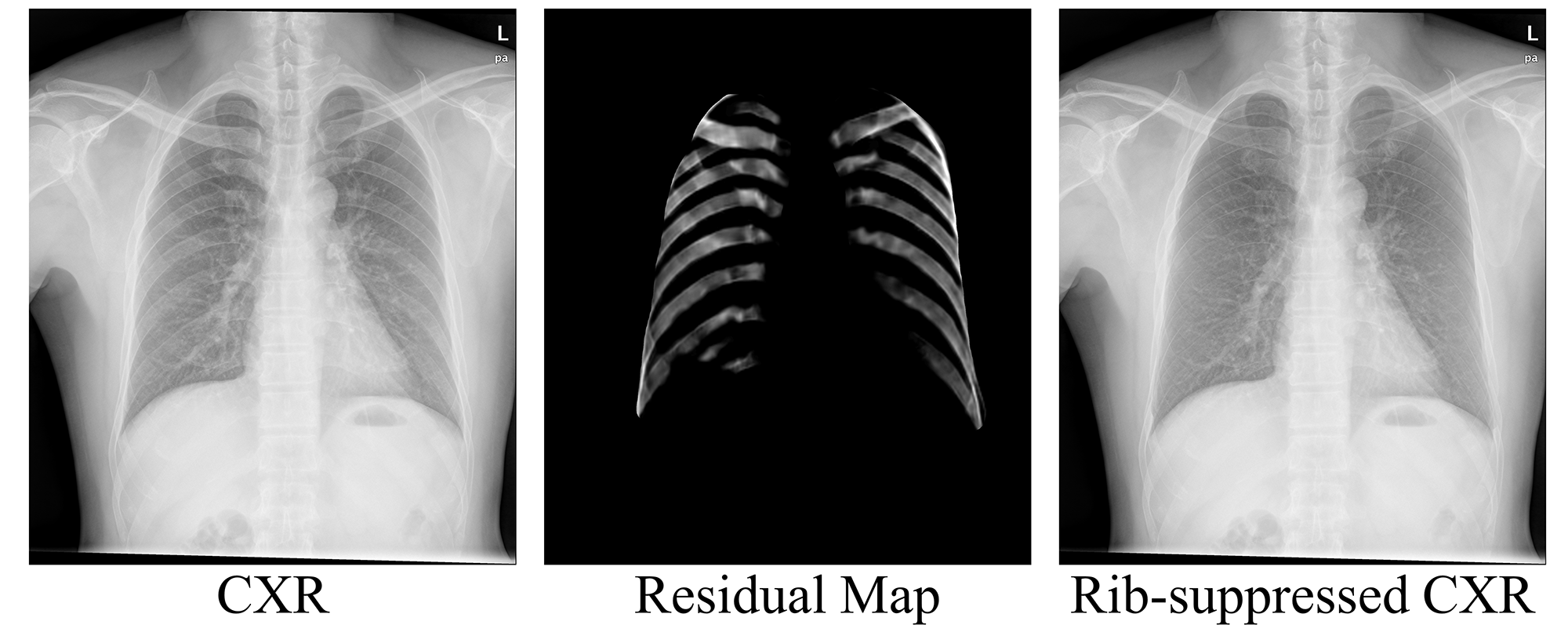}
		\caption{Rib suppression. RSGAN proposes to predict the residual map for a CXR without label. The final rib-suppressed image is obtained by subtracting it from the input CXR.}
		\label{fig:overview}
	\end{figure*}
	
	Clinical evidence indicates that bone-suppressed CXR can improve the interpretation of the CXR images and diagnostic reliability by markedly reducing the visibility of particular superimposed structures in chest radiographs~\citep{elongatedStructure,dual-energyCT}. Rib suppression is also an important preprocessing step for a computer‐aided diagnosis (CAD) system contributing to lung segmentation and tuberculosis detection~\citep{jaeger2013automatic,maduskar2013improved}. Automatically detecting and removing the rib structures in a CXR by applying image-processing techniques can simplify the feature extraction and analysis stage in a CAD system.
	Fig.~\ref{fig:overview} shows an example of rib suppression. Dual-energy (DE) CXRs can provide bone-free images, but motion artifacts are unavoidable due to cardiac motion and breath. 
	In a post-processing setting, various methods are proposed to generate bone-suppressed CXRs and can be categorized as: (1) physical model based and (2) deep learning based. 
	The physical model-based methods mimic the bone structures depending on various physical models and obtain bone-free results via subtracting bone from the original CXR.
	These approaches require manually annotated bone mask for each CXR image~\citep{elongatedStructure,von2016novel,simko2009elimination,ClavicleSeg,wu2012learning}.
	
	Learning-based methods can be divided into two subcategories according to the source of manual ground truth: (1) DE CXR based and (2) digitally reconstructed radiography (DRR) based.
	For DE CXR based learning models, the suppression of bones in CXR images is learned with a specifically designed artificial neural network~\citep{MTANN,Chen2014SeparationOB} or convolutional neural network (CNN)~\citep{MSCNNforBoneSup,BoDualEnergy,chenDualEnergy}.
	However, limited amount of DE CXR images impedes the sufficient learning of CNNs.
	DRR based learning methods attempt to utilize the structural prior knowledge in CT domain.
	Clinically, the bone component in the DRR domain paired with original chest DRR is readily accessible through projecting bone from a set of CT images, making it possible to learn a bone-suppressed model~\citep{DecGAN}.
	Although DRR appears similar to CXR, they have different contrasts due to simulation assumptions.
	Domain adaptation based on CycleGAN~\citep{CycleGAN2017} is employed to reduce the gap through transferring image from CXR domain to DRR domain.
	After that, bone component of CXR image can be generated by a model trained on large annotated DRR data with less domain gap.
	To obtain rib-suppressed results, \citet{LiTMI} subtracts bone decomposition from the original high-resolution CXR.
	By histogram matching on CXR in the area of inner-rib mask, it can produce high-resolution results with inter-rib information unchanged, which is meaningful for clinical diagnosis.
	
	\subsection{Related Work}
	Recently, GAN-based methods have achieved success in the domain adaptation. The principle of GAN is to introduce a discriminator to distinguish the generated image domain from the real image domain, while enforce the generator to trick the discriminator in order to generate an image belonging to the domain of real image.
	And the performance of domain adaptation improves during the adversarial learning between the generator and the discriminator.
	Particularly, CycleGAN~\citep{CycleGAN2017} constructs a forward- and backward-generation cycle to ensure the bilateral consistency. And this cycle-consistency constrain enables unpaired data transfer from each other~\citep{zhang2018task,kamnitsas2017unsupervised,zhang_cvpr2018}.
	However, the cycle-consistency in CycleGAN is limited to image-level, resulting in details drop at the backward generation stage.	
	
	To improve the preservation of details in performing domain adaptation, MUNIT~\citep{huang2018munit} and DRIT~\citep{DRIT} disentangle the latent space in the generator into content- and style-specific representation.
	In this way, only the domain-variant features will be exchanged and the domain-invariant feature will be well-preserved.
	Many recent medical applications, like synthesis~\citep{BenEMBC} and cross-modality segmentation~\citep{yangMICCAI}, adopt the concept of feature disentanglement to improve the performance.	
	A perfect example can be found in \citet{ChartsiasMedIA}, which improved the high-level representation on 2D medical images by disentangling latent space into spatial anatomical and non-spatial modality representation.
	Despite the success in generating more details via feature disentanglement, these methods still confront the blur issue in the generated images. It is because that the recovery capacity of the decoder is limited and it is hard to reconstruct realistic noise distribution.
	This motives us to propose an easy-to-generate output of the decoder.
	Instead of directly generating images in the target domain, our RSGAN aims at generating a residual map, which contains domain-invariant anatomical structure and residual intensity in a specific domain.
	
	The annotations of CT volume, e.g. lung and bone, can be projected to the DRR domain. By leveraging the information from DRR images, several studies developed research in X-ray decomposition~\citep{AlbarqouniDRR,DecGAN}, lung enhancement~\citep{GozesDRR,DecGAN}, and rib-suppression~\citep{DecGAN,LiTMI} tasks.
	To mitigate the domain gap between CXR and DRR, some of these methods~\citep{DecGAN,LiTMI} consist of the domain adaptation between DRR and CXR images.
	Specially, DecGAN~\citep{DecGAN} proposes a CycleGAN-based network and decomposes input into different components (bone, lung and other soft-tissue structures), then adaptively combining the decomposed components to implement the rib-suppression or lung enhancement.
	However, the rib-suppressed DRR is not strictly equal to the linear combination with the projection of lung and other soft-tissue structures. The realistic suppression indicates that the rib area in the CT volume should be replaced with intensities of soft tissue. In DecGAN~\citep{DecGAN}, this part of soft tissue is omitted and considered to be air.
	Furthermore, the CycleGAN-based framework drops details and results in a low-resolution generation. 
	\citet{LiTMI}, a further work of DecGAN~\citep{DecGAN}, attempt to generate more details in a coarse-to-fine manner.
	The generated rib-mask would be subtracted from the CXR images with ad-hoc histogram matching.
	However, the performance of rib suppression is affected by the accuracy of generated rib mask. A less accurate rib mask would result in sharp intensity changes at the rib edges in the rib-suppressed image.
	This approach has the following drawbacks: 
	\begin{enumerate}
		\item lacking structure-consistency in CXR during domain adaptation;
		\item neglecting the difference between the decomposed bone and the intensity residue resulting from a CXR image subtracted by the corresponding rib-suppressed prediction;
		\item lacking a formulation to promote a consistent rib mask, which may result in sharp changes around rib edges.
	\end{enumerate}
	
	\subsection{Contributions}
	
	To address the above challenges and obtain better rib suppression results for real CXR, 
	we propose a generative adversarial network (GAN) based disentanglement learning framework called Rib Suppression GAN (RSGAN). 
	Our contributions are three folds: 
	\begin{enumerate}
		\item We transfer the prior knowledge from the DRR domain to the CXR domain with disentangled structure- and contrast-specific generators.
		
		\item We predict residual map of ribs instead of the rib-suppressed image, enabling features to be effectively recovered from a typical decoder.
		
		\item We formulate an adaptive loss function to enhance intensity-consistency at the inter-rib regions and preserve the details overlapped by ribs.
	\end{enumerate}
	
	The remainder of this paper is organized as follows.
	In Section~\ref{sec:method}, we detail the proposed rib-suppressed method--RSGAN.
	In Section~\ref{sec:experiment}, we describe the dataset and the metrics for evaluating the performance of rib suppression.
	In Section~\ref{sec:result}, we show predicted rib-suppressed images for each competing method and the feasibility of RSGAN in downsteam applications.
	In Section~\ref{sec:discussion}, we discuss the improvement of RSGAN in detail.
	And we conclude in Section~\ref{sec:conclusion}.
	
	\begin{figure}[!htbp]
		\centering
		\includegraphics[scale=.85]{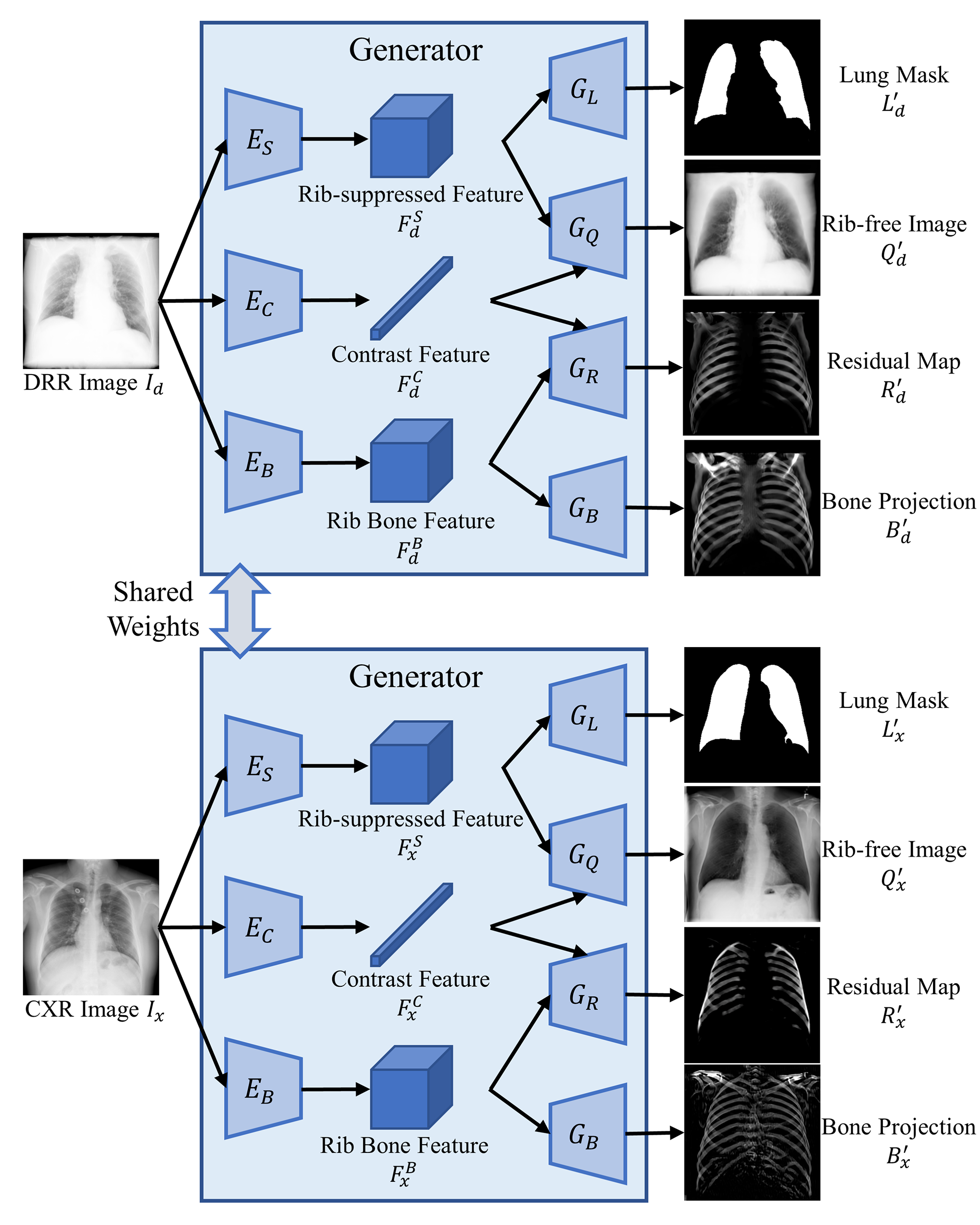}
		\caption{Structure of the disentangling generator: rib-suppressed, contrast, rib bone features are extracted from both DRR and CXR, and decoded to lung mask, rib-suppressed image, residual map, and bone projection.}
		\label{fig:method_a} 
	\end{figure}
	
	\section{Methods} \label{sec:method}
	
	\subsection{Overview}
	
	Denoting a CXR image by $I$. It is assumed that 
	\begin{equation}
		I = Q + R,
	\end{equation}
	where $Q$ is the rib-suppressed image and $R$ is the rib residual image. 
	While direct approaches explicitly predicts the rib-suppressed image $Q$ from $I$, we attempt to estimate $R$ from $I$ instead and then subtract it from $I$ to finally get $Q$.
	
	We propose a GAN-based disentanglement learning framework RSGAN to suppress rib bones in CXR images by leveraging anatomical knowledge from unpaired CT/DRR images.
	Although the contrast is different between CXR and DRR images, the anatomical structures are similar in both images.
	Based on this fact, our framework disentangles structure and contrast with separate generators for both CXR and DRR images. 
	The rib bone features learned from the DRR domain are then combined with the contrast feature of the CXR domain to construct the CXR-based residual map.
	The final rib-suppressed CXR is achieved through subtracting transferred ribs from the input CXR image as shown in Fig.~\ref{fig:overview}.
	To facilitate to the description of the proposed approach, the mathematical setting and notations are listed in Table~\ref{tab:notation}.

	\begin{table*}[!t]
		\caption{\label{tab:notation}The mathematical setting and notations used in this paper.}
		\centering
		\begin{tabular}{c|l}
			\hline
			$I_x$ & CXR image \\
			\hline
			$I_d$ & DRR image \\
			\hline
			$Q_d$ & Rib-suppressed image of DRR \\
			\hline
			$R_d$ & Residual map of DRR, where $R_d = I_d - Q_d$ \\
			\hline
			$B_d$ & Bone component projection of DRR \\
			\hline
			$L_d$ & Lung mask of DRR \\
			\hline
			$F^C_x$, $F^C_d$ & Contrast feature extracted by encoder $E_C$ for CXR and DRR \\
			\hline
			$F^B_x$, $F^B_d$ & Rib bone feature extracted by encoder $E_B$ for CXR and DRR \\
			\hline
			$F^S_x$, $F^S_d$ & Rib-suppressed feature extracted by encoder $E_S$ for CXR and DRR \\
			\hline
			$Q_x'$, $Q_d'$ & Rib-suppressed image generated by decoder $G_Q$ for CXR and DRR \\
			\hline
			$R_x'$, $R_d'$ & Residual map generated by decoder $G_R$ for CXR and DRR \\
			\hline
			$B_x'$, $B_d'$ & Bone component projection generated by decoder $G_B$ for CXR and DRR \\
			\hline
			$L_x'$, $L_d'$ & Lung mask generated by decoder $G_L$ for CXR and DRR \\
			\hline
			$I_x'$, $I_d'$ & Reconstructed CXR and DRR, where $I_x'=Q_x'+R_x'$ and $I_d'=Q_d'+R_d'$ \\
			\hline
			$I_x''$, $I_d''$ & Cycle reconstructed CXR and DRR \\
			\hline
			$I'_{x\rightarrow d}$, $I'_{d\rightarrow x}$ & Domain transferred image between CXR and DRR \\
			\hline
			$Q'_{d\rightarrow x}$ & Predicted rib-suppressed CXR transferred from DRR, where $Q'_{d\rightarrow x} = G_Q(E_C(I_x), E_S(I_d))$ \\
			\hline
			$M'_x$ & Bone mask for CXR generated by bilateral filter and threshold segmentation, where $M'_x = [F_{bilateral}(B_x')>\theta_{thresh}]$ \\
			\hline
		\end{tabular}
	\end{table*}
	
	\subsection{Disentangled representation in generator}
	Fig.~\ref{fig:method_a} shows the disentangled representations in our generative model.
	In the generator, each input is disentangled into three components, including anatomical content, image contrast, and rib bones, with three encoders $E_{S}$, $E_{C}$, and $E_{B}$, respectively. The output has four predictions (lung mask, rib-suppressed image, residual map of rib, and bone component projection) from four decoders $G_{L}$, $G_{Q}$, $G_{R}$, and $G_{B}$, respectively.
	
	In the decoding path, $G_Q$ inputs with contrast and rib-suppressed features and outputs the rib-suppressed image.
	Improving from \citet{DecGAN}, we inpaint the rib-removed area in CT volume based on its surrounding tissue intensities to simulate more real suppression performance, which projected as the ground truth for rib-suppressed DRR.
	$G_R$ takes the contrast and rib bone feature as inputs, outputs a residual map.
	The residual map is utilized to generate a rib-suppressed image with more details by subtracting it from the input image.
	The input image can be reconstructed by adding the rib-suppressed image and the residual map.
	$G_B$ only takes the rib bone feature as input and outputs the bone component projection.
	The ground truth of the bone component projection for the DRR image is generated by projecting a separate CT rib bone component obtained in a 3D volume.
	By generating bone projection, better rib bone features can be obtained for residual map prediction, and a bone mask is available due to the clearer rib edge than the residual map.
	$G_L$ inputs rib-suppressed features and outputs the lung mask, which can help the model pay more attention to the lung area.
	
	For CXR images, it is difficult to achieve the desirable rib-suppressed results due to the overlap of multiple tissues (e.g., ribs, clavicles, lung). 	
	To relieve the interference from other organs, the specific representation of ribs should be well constructed.
	CT scans provide the 3D structure of the ribs that are without any overlap and hence easily annotated. Thus tailored rib-suppressed DRR images can be delivered by projecting from the corresponding rib-suppressed CT slices. The rib features of CXR images then can be transferred from those of DRR images.
	As illustrate in Fig.~\ref{fig:method_a}, the mapping mode of rib-suppression is learned from the DRR domain and supervised from three aspects, including rib-suppression, residual map of ribs, and bone component.
	The supervised loss is given as,		
	\begin{equation}
		\label{eq:superviseRecLoss}
		\begin{aligned}
			\mathcal{L}_{su}& =\left \| Q_d'-Q_d \right \|_1 + \left \| R_d'-R_d \right \|_1 
			+ \left \| B_d'-B_d \right \|_1,
		\end{aligned}
	\end{equation}	
	where $\left \| \cdot \right \|_1$ is a $L_1$ loss,	$Q_d$ and $Q_d'$ are the ground truth and prediction of rib-suppressed DRR image, respectively, $R_d$ and $R_d'$ are the ground truth and predicted residual map of bone in DRR image, respectively, and $B_d$ and $B_d'$ are the ground truth and prediction of bone component projection in DRR image, respectively.
	
	\begin{figure*}[!htbp]
		\centering
		\includegraphics[scale=.9]{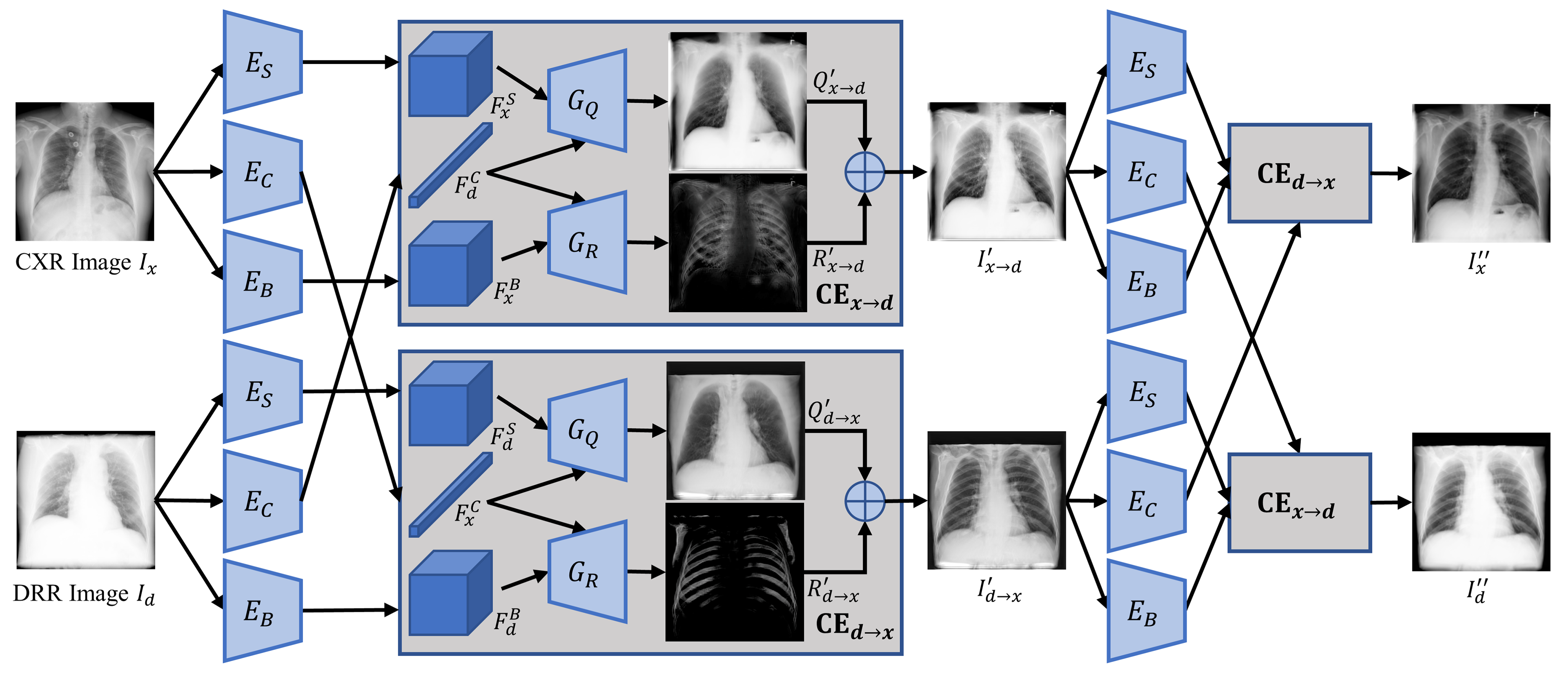}
		\caption{Flowchart of domain adaptation via contrast exchanging (CE) block, the input image is transformed to the target domain by the first CE block and back to the input domain by the second one.}
		\label{fig:method_b} 
	\end{figure*}
	
	\subsection{CXR disentanglement via domain adaptation}
	Directly disentangling ribs from CXR under the supervision of the annotated rib in DRR is ineffective due to the domain gap. 
	To reduce the impact of domain variance in rib-suppression, our RSGAN, shown in Fig.~\ref{fig:method_b}, employs cycle-consistency learning to implement CXR disentanglement via domain adaption.
	Particularly, we separate the learning into domain-invariant and domain-specific features.
	Here, we assume that the rib bone features and the rib-suppressed features disentangled from CXR images or DRR images are domain-invariant, while the contrast features are domain-specific.
	In this way, a domain-specific image can be generated by exchanging the contrast feature from the target domain via contrast exchanging (CE) block.
	
	The requirement of transferring the rib-suppressed map from the DRR domain to the CXR domain is to achieve the specific style of these two domains.
	Here, two discriminators are involved to evaluate the generations for DRR style and CXR style, respectively.
	Additionally, the third discriminator is utilized to assess the predicted difference of bone projection from CXR and DRR images, so that to sufficiently and precisely transfer the rib structures from DRR to CXR.
	Adversarial learning is employed to minimize the domain distance between real image and prediction, and adversarial losses of the generator and discriminator are defined as,
	\begin{equation}
		\label{eq:D_adv-stage1}
		\begin{aligned}
			\mathcal{L}_{D_{adv}}& =\mathbb{E}_{I_x}(D_x(I_x)-1)^2+\mathbb{E}_{I'_{d\rightarrow x}}(D_x(I'_{d\rightarrow x}))^2 \\
			& + \mathbb{E}_{I_d}(D_d(I_d)-1)^2+\mathbb{E}_{I'_{x\rightarrow d}}(D_d(I'_{x\rightarrow d}))^2 \\
			& + \mathbb{E}_{B_d'}(D_B(B_d')-1)^2+\mathbb{E}_{B_x'}(D_B(B_x'))^2,\\
			\mathcal{L}_{G_{adv}}& =\mathbb{E}_{I'_{d\rightarrow x}}(D_x(I'_{d\rightarrow x})-1)^2 + \mathbb{E}_{I'_{x\rightarrow d}}(D_d(I'_{x\rightarrow d})-1)^2 \\
			& + \mathbb{E}_{B_x'}(D_B(B_x')-1)^2,
		\end{aligned}
	\end{equation}  
	where $I_x$ and $I_d$ denote the CXR and DRR image.
	$I'_{d\rightarrow x}$ denotes the predicted CXR image transferred from DRR domain.
	$I'_{x\rightarrow d}$ denotes the predicted DRR image transferred from CXR domain.
	$B_d'$ and $B_x'$ refer to the prediction of bone component projection in DRR image and CXR image, respectively.
	Two discriminators $D_x$ and $D_d$ are leveraged to evaluate the performance of the reconstructed domain-transfer images in CXR domain and DRR domain, respectively, which retain anatomical structure but with exchanged contrast.
	$D_B$ refers the bone component discriminator to distinguish the predicted bone projection from CXR to DRR.
	
	To improve transferability between two domains, we constrain learning on both feature- and image-level reconstructions.
	At the feature level, we employ contrast- and structure-consistency losses. 
	Contrast-consistency loss $\mathcal{L}_{c}$ is used to keep the contrast feature identical within the same domain.
	The structure-consistency loss $\mathcal{L}_{s}$ is used to protect the structures from deterioration when the contrast changes.
	$\mathcal{L}_{c}$ and $\mathcal{L}_{s}$ are defined as follows:	
	\begin{equation}
		\mathcal{L}_{c}=\left \| E_C(I_x)-E_C(I'_{d\rightarrow x}) \right \|_1 + \left \| E_C(I_d)-E_C(I'_{x\rightarrow d}) \right \|_1,
	\end{equation}	
	\begin{equation}
		\begin{aligned}
			\mathcal{L}_{s}& =\left \| E_S(I_x)-E_S(I'_{x\rightarrow d}) \right \|_1 + \left \| E_S(I_d)-E_S(I'_{d\rightarrow x}) \right \|_1 \\
			& + \left \| E_B(I_x)-E_B(I'_{x\rightarrow d}) \right \|_1 + \left \| E_B(I_d)-E_B(I'_{d\rightarrow x}) \right \|_1,
		\end{aligned}
	\end{equation}	
	where $E_C$ refers to the contrast encoder,	$E_S$ refers to the rib-suppressed encoder, $E_B$ refers to the rib bone encoder, 
	$I_{x\rightarrow d}$ means transferring the CXR image to the DRR domain, and $I_{d\rightarrow x}$ denotes the reverse transfer.

	To ensure the generation be consistent with the real images, a pixel-wise $L_{1}$ constraint is used to yield a reconstructed image close to the real one at the image-level. 
	Meanwhile, a cycle-consistency $L_{1}$ constraint is introduced to ensure the pixel similarity between an original input and the corresponding cycle-transferred generation as shown in Fig.~\ref{fig:method_b}.
	The pixel-wise reconstruction loss $\mathcal{L}_{rec}$ and the cycle-consistency loss $\mathcal{L}_{cyc}$ are utilized to ensure the image-to-image translation, which are written as follows,
	\begin{equation}
		\begin{aligned}
			\mathcal{L}_{rec} =\left \| I_x'-I_x \right \|_1 + \left \| I_d'-I_d \right \|_1,
		\end{aligned}
	\end{equation}	
	\begin{equation}
		\begin{aligned}
			\mathcal{L}_{cyc} = \left \|  I_x''-I_x \right \|_1 + \left \|  I_d''-I_d \right \|_1,
		\end{aligned}
	\end{equation}	
	where $I_x'$ and $I_d'$ refer to the forward prediction of $I_x$ and $I_d$, respectively, and $I_x''$ and $I_d''$ denote the cycle predictions, respectively.
	$I_x'$ and $I_x''$ are formulated as
	\begin{equation}
		\begin{aligned}
			I_x'&=G_Q(E_C(I_x),E_S(I_x))+G_R(E_C(I_x),E_B(I_x)), \\
			I_x''&=G_Q(E_C(I'_{d\rightarrow x}),E_S(I'_{x\rightarrow d}))\\&+G_R(E_C(I'_{d\rightarrow x}),E_B(I'_{x\rightarrow d})),
		\end{aligned}
	\end{equation}
	$I_d'$ and $I_d''$ are similarly formulated.
	
	
	
	\begin{figure*}[!htbp]
		\centering
		\includegraphics[scale=.9]{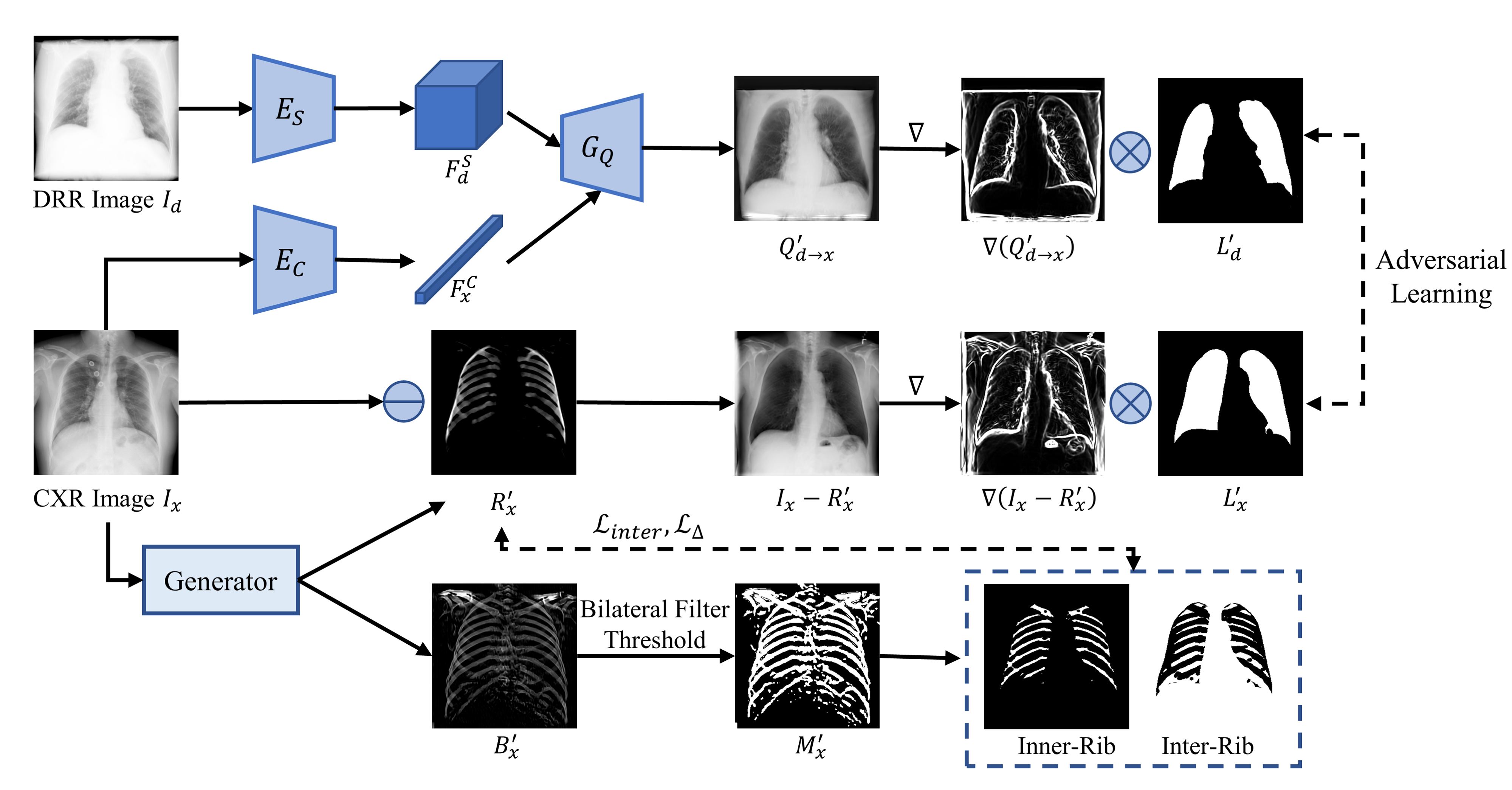}
		\caption{Rib-suppressed adaptive loss functions used in our proposed RSGAN, consist of adversarial learning on the gradient map of rib-suppressed image, and inner-rib and inter-rib constrains.}
		\label{fig:method_c} 
	\end{figure*}
	
	\subsection{Rib-suppression learning on CXR}
	The above domain adaptation learning is built up using the global intensity distribution.
	The translation of local details is ignored.
	Hence, as shown in Fig.~\ref{fig:method_c}, additional local constraints are applied on the rib-suppression learning on CXR images from three aspects: (1) handling rib residue; (2) inter-rib intensity consistency; and (3) inner-rib detail preservation.
	
	\paragraph{Handling rib residue}
	Ribs may reside in the rib-suppression prediction during the domain adaptation learning. 
	Here, we employ a gradient-based constraint within the lung regions to eliminate the influence of intensity various across images and enhance smooth translation at the rib regions for a rib-suppression prediction. 
	Gradient supervision is applied on predicted rib-suppression of CXR images and the corresponding residual maps of ribs during the adversarial learning, which is defined as,	
	\begin{equation}
		\label{eq:D_gradient}
		\begin{aligned}
			\mathcal{L}_{D_{\bigtriangledown}}& = \mathbb{E}_{Q'_{d\rightarrow x}}(D_{\bigtriangledown}(\bigtriangledown Q'_{d\rightarrow x}\cdot L'_d)-1)^2 \\ & + \mathbb{E}_{R'_x}(D_{\bigtriangledown}(\bigtriangledown(I_x-R'_x)\cdot L'_x))^2,\\
			\mathcal{L}_{G_{\bigtriangledown}}&=\mathbb{E}_{R'_x}(D_{\bigtriangledown}(\bigtriangledown(I_x-R'_x)\cdot L'_x)-1)^2,
		\end{aligned}
	\end{equation}
	where $\bigtriangledown$ refers to the gradient operator, $D_{\bigtriangledown}$ refers to a discriminator aiming to regularize the gradient map of rib-suppressed image for a smooth transition on rib edges, $Q'_{d\rightarrow x}$ is the predicted rib-suppressed CXR images transferred from the DRR domain, $R'_x$ denotes the predicted residual map of ribs in CXR and $L'_x$ and $L'_d$ refer to the predicted lung mask in CXR and DRR image, respectively.
	The lung mask decoder is trained with a Dice loss on a set of annotated CXR images and is frozen afterward.
	
	\paragraph{Inter-rib intensity consistency}
	Inter-rib intensities should be consistent between the input CXR image and the corresponding rib-suppressed prediction.
	The intensity-consistency loss is defined as:	
	\begin{equation}
		\label{eq:inter}
		\mathcal{L}_{inter} = \left \| R'_x\cdot [(1-M'_x)\cup(1-L'_x)] \right \|_1,
	\end{equation}	
	where $M'_x$ denotes the corresponding bone mask of $I_x$, which is generated from $B'_x$ by utilizing bilateral filter and threshold segmentation, and $\cup$ denotes the union region of two input masks.
	Note that, the bone mask $M'_x$ can not be generated from the residual map $R'_x$. Because it will bring unstable texture changes to the residual map when introducing $\mathcal{L}_{G_{\bigtriangledown}}$. The bone mask $M'_x$ will be influenced by the disturbance of the residual map and enlarge the bias during the training stage.
	Instead, the bone component projection $B'_x$ is more stable and stops the gradient from $\mathcal{L}_{G_{\bigtriangledown}}$, which is necessary to generate the bone mask.
	
	\paragraph{Inner-rib detail preservation}
	To preserve the trachea overlapped with ribs in the predicted residual map of ribs, we employed a Laplacian-based loss $\mathcal{L}_{\bigtriangleup}$ to eliminate the overlapped regions in the residual prediction.
	Laplacian as a regularization method encourages generating smoother residual maps~\citep{LiTMI}.
	$\mathcal{L}_{\bigtriangleup}$ is defined as,	
	\begin{equation}
		\label{eq:inner}
		\mathcal{L}_{\bigtriangleup} = \left \| \bigtriangleup R'_x\cdot (M'_x\cup L'_x) \right \|_1,
	\end{equation}
	where $\bigtriangleup$ represents the Laplacian operator.
	
	\subsection{Total loss function} \label{sec:method_init}
	The above extra adaptive loss functions are employed to refine the performance of rib suppression on local details, which only work on the fine-tuning stage and need to be removed in the initial stage.
	The total loss function of the generative model in the initial stage and in the fine-tuning stage can be summarized as,
	\begin{equation}
		\label{eq:stage1}
		\begin{aligned}
			\mathcal{L}_{D_{init}}&=\lambda_{adv}\mathcal{L}_{D_{adv}},\\
			\mathcal{L}_{G_{init}}& =\mathcal{L}_{su}+\lambda_{adv}\mathcal{L}_{G_{adv}}\\
			&+ \lambda_{f}(\mathcal{L}_{c}+\mathcal{L}_{s})+\lambda_{i}(\mathcal{L}_{rec}+\mathcal{L}_{cyc}),
		\end{aligned}
	\end{equation}
	\begin{equation}
		\label{eq:stage2}
		\begin{aligned}
			\mathcal{L}_{D_{fine}}&=\lambda_{adv}\mathcal{L}_{D_{adv}}+\lambda_{G_{\bigtriangledown}}\mathcal{L}_{{D}_{\bigtriangledown}},\\
			\mathcal{L}_{G_{fine}}& =\mathcal{L}_{su}+\lambda_{adv}\mathcal{L}_{G_{adv}}\\
			&+ \lambda_{f}(\mathcal{L}_{c}+\mathcal{L}_{s})+\lambda_{i}(\mathcal{L}_{rec}+\mathcal{L}_{cyc})\\ 
			&+\lambda_{G_{\bigtriangledown}}\mathcal{L}_{G_{\bigtriangledown}} + \lambda_{inter}\mathcal{L}_{inter} + \lambda_{\bigtriangleup}\mathcal{L}_{\bigtriangleup},
		\end{aligned}
	\end{equation}	
	where we set $\lambda_{adv}=1$, $\lambda_{f}=1$, $\lambda_{i}=10$, $\lambda_{G_{\bigtriangledown}}=10$, $\lambda_{inter}=500$, and $\lambda_{\bigtriangleup}=1$ with experimental experience.
	
	\section{Experimental Settings} \label{sec:experiment}
	
	\subsection{Data}
	We utilize two CT datasets and four CXR datasets in our experiment.
	CT datasets involve 896 CT volumes from LIDC-IDRI~\citep{LIDC-IDRI} and 777 CT volumes from 2017~\footnote{https://tianchi.aliyun.com/competition/entrance/231601/introduction} and 2019~\footnote{https://tianchi.aliyun.com/competition/entrance/231724/introduction} TianChi AI Competition for Healthcare
	organized by Alibaba and all the CT volumes are selected with a thickness less than 2.5 mm.
	CXR datasets involve 11,200 CXRs of size $512\times512$ from TBX11K~\citep{liu2020rethinking}, 112,120 CXRs of size $1024\times1024$ from chest-14~\citep{ChestX-Ray8}, 138 CXRs of size around $4020\times4892$ from Montgomery County (MC)~\citep{MC&shenzhen}, and 662 CXRs of size around $3000\times3000$ from Shenzhen Hospital~\citep{MC&shenzhen}.
	As for the CT dataset, we train a U-Net model for rib segmentation on 199 CT volume selected from LIDC-IDRI with manual labels and generate rib masks for the whole CT dataset.
	Then, based on the rib masks, we suppress the region of ribs on CT volumes with an inpainting method~\citep{cv2.inpaint}.
	To generate the DRRs, we utilize DeepDRR~\citep{DeepDRR2019} to project each CT volume into 42 projection slices which are uniformly sampled from $-10^{\circ}$ to $10^{\circ}$ in azimuth and elevation angles, respectively.
	Similarly, we generate the projection of rib-suppressed volume, rib region, and lung area with the same projection strategy.
	All the DRRs are resized to $320\times320$ based on bilinear interpolation for training.
	As for the CXR dataset, all the CXRs are resized to $320\times320$ when input into the network and back to their original resolution for rib suppression.
	We utilize the training set of TBX11K, Chest-14, and MontgomerySet for training, and others for testing.
	
	\begin{figure}[!htbp]
		\centering
		\includegraphics[scale=.9]{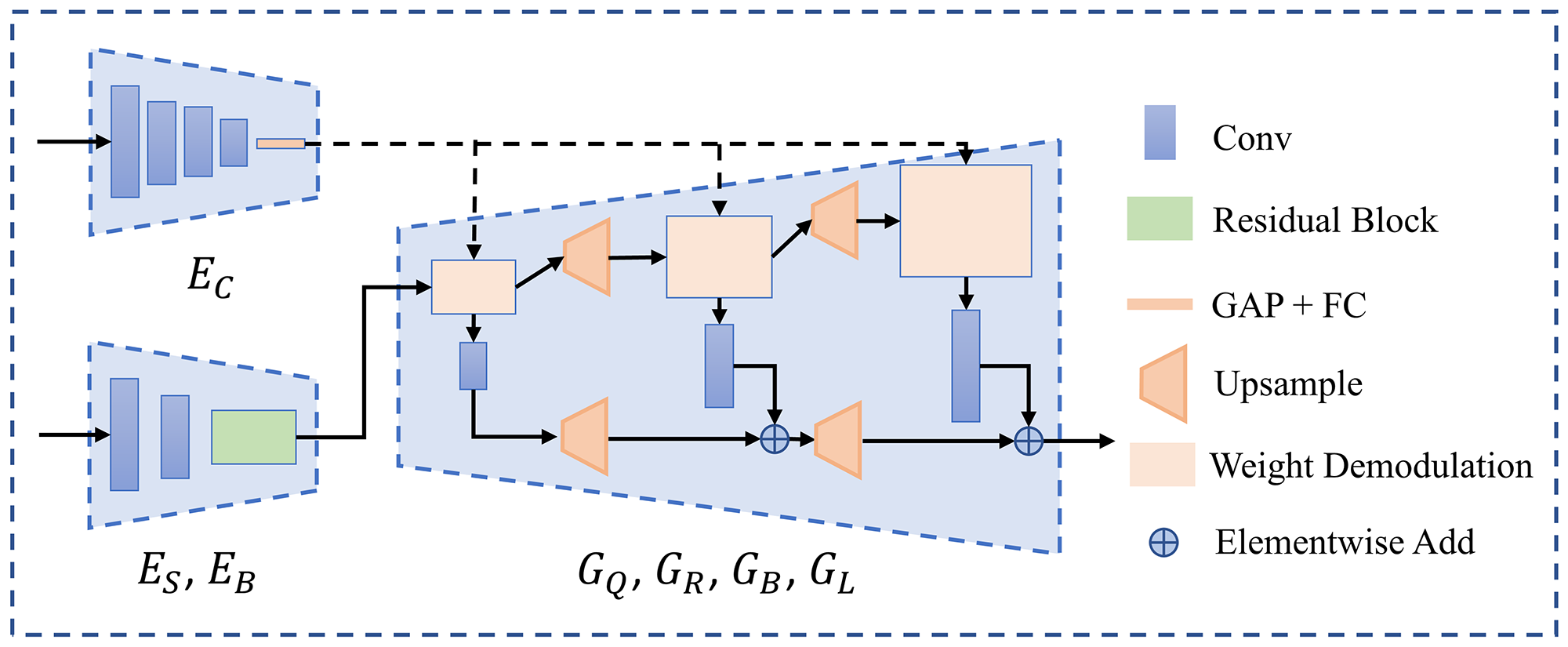}
		\caption{Implementation details of blocks in the generator.}
		\label{fig:generator}
	\end{figure}
	
	\subsection{Implementation details}
	
	Fig.~\ref{fig:generator} illustrates the details of three kind of blocks in the generator.
	The contrast encoder $E_C$ is consisted of four convolutional layers with a kernel size of $3\times3$ and a stride of 2, followed with a global average pooling (GAP) and a fully connection layer of 256 output channels.
	The rib-suppressed encoder $E_{S}$ and the rib bone encoder $E_B$ have the same structure, consisting of two convolutional layers with a kernel size of $3\times3$ and a stride of 2, followed with a residual block of 256 output channels.
	In this paper, the size of features encoded from $E_{S}$ and $E_B$ is set to be $80\times80$ with the channel of 256.
	The rib-suppressed image decoder $G_Q$ and the residual map decoder $G_R$ have the same structure as shown in Fig.~\ref{fig:generator}, involving three weight demodulation blocks~\citep{Karras2019stylegan2}, and each weight demodulation block is followed with an upsampling layer to pass a higher dimension of feature to the next block.
	The scale factor of the upsampling layer is 2.
	Three different resolutions of feature maps in the decoder are set to be $80\times80$, $160\times160$, and $320\times320$.
	Every block includes another sub-branch of convolutional layer to generate an output image in a higher resolution.
	The lower resolution image is merged with the higher resolution one outputed by next block via an upsampling layer and an element-wise addition.
	The contrast feature is fed into each weight demodulation block as shown by the dotted arrow.
	The bone component decoder $G_B$ has similar structure with $G_Q$ and $G_R$, replacing the demodulation block with residual block, and the final output is followed with a Tanh activation.
	The lung mask decoder $G_L$ is the same as $G_B$, but outputs with a Sigmoid activation to predict the mask of lung.
	Note it, the contrast feature is not fed into $G_B$ and $G_L$.
	
	Four discriminators are utilized in our proposed RSGAN. All discriminators are built with PatchGAN~\citep{CycleGAN2017}, consisting of four convolutional layers with a kernel size of 4 and a stride of 2 and one convolutional layer with a kernel size of 1.
	
	The proposed network is developed with PyTorch and trained on a GeForce RTX 2080 Ti GPU.
	All the networks are trained using Adam optimizer with a learning rate of $1\times10^{-5}$.
	We train the proposed network with 40,000 iterations in the initial stage, and 10,000 iterations for the fine-tuning stage with a batch size of 1.
	
	\subsection{Evaluation metrics}
	Four metrics are utilized to evaluate rib suppression quality, including Weber Contrast~\citep{weber}, Learned Perceptual Image Patch Similarity (LPIPS)~\citep{zhang2018unreasonable}, Peak Signal Noise Rate (PSNR), and Structural Similarity Index Measure (SSIM).
	
	Weber Contrast~\citep{weber} provides an estimation of rib-suppression performance on the boundaries for CXR images by calculating the contrast gap between the rib-suppressed region and the background. It is defined as 
	$C_w= I_r/I_b~-~1$,	
	where $I_r$ denotes the average intensity of the manually annotated rib region.
	$I_b$ denotes the confined background, corresponding to the surrounded region of ribs in a radius of fewer than 5 pixels.
	Lower Weber Contrast means higher similarity between the rib-suppressed region and the background, suggesting better visualization of the predicted rib-suppressed CXR.
	
	LPIPS~\citep{zhang2018unreasonable} is a metric to measure the perceptual similarity at feature level, enabling to evaluate the difference between the generated and real images at multiple feature space. LPIPS is defined as follow,
	\begin{equation}
		d(x,x_0) = \sum_{l}\frac{1}{H_lW_l}\sum_{h,w}\left \| \omega_l\odot(\hat{y}^l_{hw}-\hat{y}^l_{0hw}) \right \|^2_2,
	\end{equation}
	where $x$, $x_0$ denote the reference and generated patch, respectively. $\hat{y}^l$, $\hat{y}^l_{0}\in\mathbb{R}^{H_l\times W_l\times C_l}$ correspond to the unit-normalized features at $l$\textit{th} layer in a feature-extraction network when inputting $x$, $x_0$.
	$\omega_l\in\mathbb{R}^{C_l}$ represents a  weighted vector to control different channels, $\odot$ refers to a channel-wise multiplication, and $\left \| \cdot \right \|_2$ is a $L_2$ loss. Note that, AlexNet~\citep{krizhevsky2012imagenet} is chosen as the feature-extraction network, and all weights in vector $\omega_l$ are set to $1$.
	
	Besides, PSNR and SSIM, the widely used image-level similarity metrics, are employed to estimate the details remained in the rib-suppressed prediction.
	PSNR presents the total pixel errors, and SSIM quantifies the structure similarity based on brightness, contrast, and structures.
	All the above-mentioned three similarity metrics are calculated over the manually labeled lung regions without bones.
	Lower LPIPS, higher PNSR and SSIM refer to a better prediction.
	
	Other two metrics are introduced to evaluate the quality of two key components in RSGAN -- residual map mechanism and domain adaptation, including Mean Absolute Error for reconstruction consistency ($\mathbf{MAE}_{rec}$) and Mean Absolute Error for cycle consistency ($\mathbf{MAE}_{cyc}$). They are defined as $\mathbf{MAE}_{rec}=\left \| I'_x-I_x \right \|_1$ and $\mathbf{MAE}_{cyc}=\left \| I''_x-I_x \right \|_1$. Lower $\mathbf{MAE}_{rec}$ and $\mathbf{MAE}_{cyc}$ refer to a better performance of the generator.
	
	\begin{figure*}[!htbp]
		\centering
		\includegraphics[scale=.9]{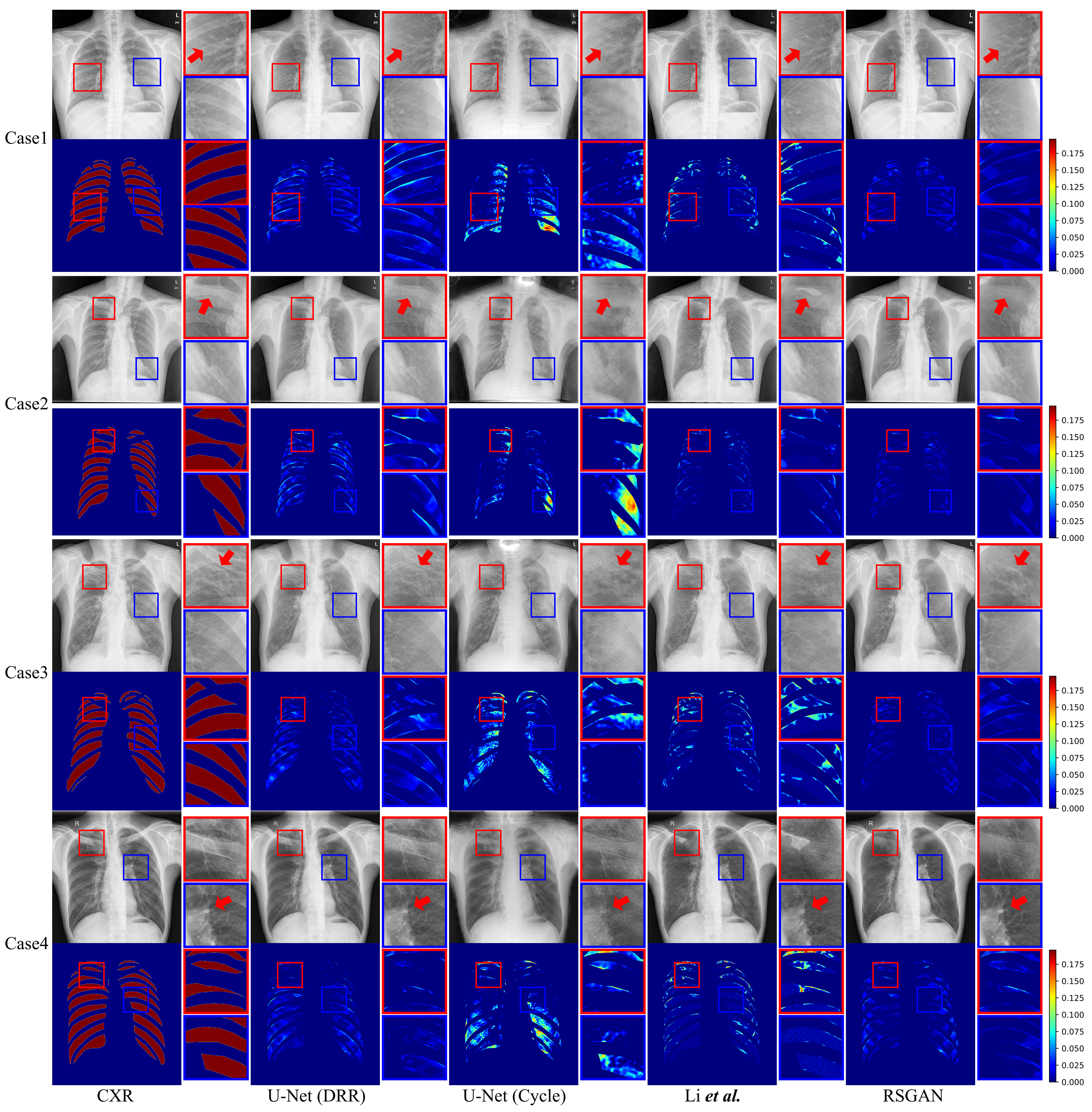}
		\caption{Rib-suppressed results for alternative methods. Case 1 and Case 2 are normal CXR images. Case 3 and Case 4 are with pulmonary tuberculosis. The first line of each case is the CXR images, and the second line is the colormap for the difference in the inter-rib area between the rib-suppressed image and the original CXR. The red region in the CXR colormap indicates the manually labeled mask of the inter-rib area. Rectangles in different colors refer to different areas in the image, and the areas are magnified at the right side of the image. Red arrows indicate some notable differences in the local area.}
		\label{fig:result}
	\end{figure*}
	
	\section{Results} \label{sec:result}
	\subsection{Alternative comparisons}
	We compare our method with U-Net (DRR), U-Net (Cycle), and \citet{LiTMI} with the above-mentioned metrics.
	For the setting of U-Net (DRR), a U-Net~\citep{unet} is trained on annotated DRR images and directly used to predict the residual map of ribs for CXR.
	For U-Net (Cycle), the CXR image is first translated to the DRR domain by CycleGAN~\citep{CycleGAN2017} and fed back to the CXR domain after rib suppression by U-Net~\citep{unet}.
	The comparisons are based on 192 CXRs with manually annotated rib and lung region, in which 99 CXRs are randomly selected from the Shenzhen hospital dataset~\citep{MC&shenzhen} and 93 come from Chest-xray14 dataset~\citep{ChestX-Ray8}.
	
	Table~\ref{tab:SuppressionQuality} and Fig.~\ref{fig:result} show the quantitative and visual results of alternative methods, respectively. RSGAN achieves state-of-the-art performance with the lowest Weber Contrast of $1.96\times10^{-2}$, LPIPS of $0.62\times10^{-2}$ and the highest PSNR of 34.62, SSIM of 0.996.
	The LPIPS, PSNR and SSIM results of RSGAN are much better than those of U-Net (DRR) and U-Net (Cycle).
	This is probably because RSGAN poses a constraint on the inter-rib area and ensures the region between ribs unchanged.
	The proposed RSGAN achieves slightly better Weber Contrast, PSNR and SSIM than those of \citet{LiTMI}, but obtains a $0.56\times10^{-2}$ drop of LPIPS. It is indicated that RSGAN introduces better perceptual similarity on the edge of the rib area. 
	As illustrated in Fig.~\ref{fig:result}, we randomly choose four CXRs including two normal and two pulmonary tuberculosis cases from Shenzhen hospital dataset~\citep{MC&shenzhen}.
	The rib-suppressed results by RSGAN have cleaner visualization in both overview and details than other methods and also have a lower difference of inter-rib area in the colormap.
	In the blue rectangle of case 1, U-Net (DRR) and U-Net (Cycle) can not suppress the ribs completely and details are smoothed out. Due to the invalidation of histogram matching, a sharp difference appears in the result of \citet{LiTMI}. Among all the methods, RSGAN removes the most rib residue and retains the richest details, which proves that RSGAN can obtain a better and robust rib-suppressed image via the disentanglement approach.
	As for the details remaining in the rib-suppressed results, the proposed RSGAN could make the detail much clearer than other methods and retain both big nidus (tuberculosis in red rectangles) and very small objects (small vascular and trachea, e.g. in blue rectangles), but \citet{LiTMI} leverage a large residual of clavicle because of bad rib prediction.
	All these advantages contribute to reducing the reading difficulty and the chance of misdiagnosis.
	
	\subsection{Ablation study}
	We investigate the effectiveness of two key components in our RSGAN: (i) residual map mechanism and domain adaptation (RMDA); and (ii) different extra adaptive loss functions.
	Residual map mechanism is accompanied with the domain adaption, and we take them as an indivisible part. The experiments with and without RMDA are performed to demonstrate the effectiveness of component RMDA.
	The detailed experimental settings are RSGAN not using residual map (nRM), RSGAN using residual map (RM) , and RSGAN using RMDA.
	In the experiment of RSGAN (nRM), the finally rib-suppressed image is directly predicted from the generator $G_Q$ in Fig.~\ref{fig:method_a}. And the generator $G_{Q}$ is trained with the following loss function,
	
	\begin{equation}
		\label{eq:rsgan(nores)}
		\begin{aligned}
			\mathcal{L}_{G_{nRM}}=\left \| Q_d'-Q_d \right \|_1,
		\end{aligned}
	\end{equation}
	where $Q_d$ and $Q_d'$ are the ground truth and prediction of rib-suppressed DRR image, respectively.
	
	In the experiment of RSGAN (RM), all the generators are preserved without the discriminators. The generators with residual map mechanism are trained with the following loss function,
	
	\begin{equation}
		\label{eq:rsgan(su)}
		\begin{aligned}
			\mathcal{L}_{G_{RM}}=\mathcal{L}_{su} + \lambda_{i}\mathcal{L}_{rec},
		\end{aligned}
	\end{equation}
	where $\lambda_{i}$ is set to $10$, which is identical with RSGAN.
	
	For RSGAN (RMDA), all the generators and discriminators are available, and all the setting are the same as the proposed RAGAN expect using the extra adaptive loss functions. 
	RSGAN (RMDA) is equal to the RSGAN only trained after the initial stage mentioned in Section~\ref{sec:method_init}, whose loss function is written as Eq.~\ref{eq:stage1}.
	
	Then we set RSGAN (RMDA) as the baseline model, and provide an ablation study on the effect of proposed extra loss functions, including rib-suppressed adversarial loss $\mathcal{L}_{G_{\bigtriangledown}}$, background consistency loss $\mathcal{L}_{inter}$, and residual smooth loss $\mathcal{L}_{\bigtriangleup}$.
	We compare the quantitative results with adding the above components in turn, and evaluate with the metrics of Weber Contrast, LPIPS, PSNR,  and SSIM. 
	
\begin{figure*}[!htbp]
		\centering
		\includegraphics[scale=.9]{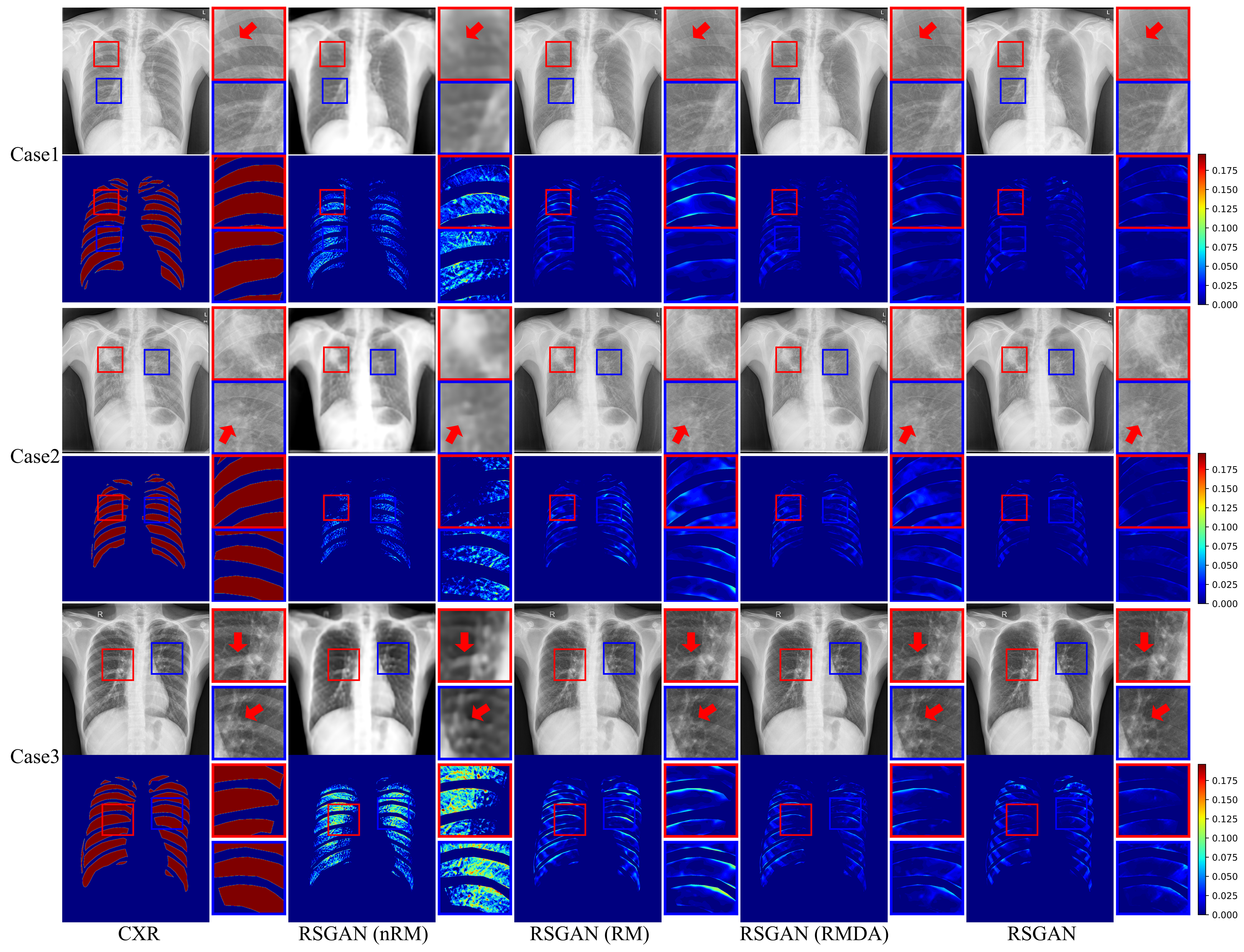}
		\caption{Rib-suppressed results for ablation study of residual map mechanism and domain adaptation. The first line of each case is the CXR images, and the second line is the colormap for the difference in the inter-rib area between the rib-suppressed image and the original CXR. The red region in the CXR colormap indicates the manually labeled mask of the inter-rib area. Rectangles in different colors refer to different areas in the image, and the areas are magnified at the right side of the image. Red arrows indicate some notable differences in the local area.}
		\label{fig:ablation_structure}
	\end{figure*}

	Table~\ref{tab:SuppressionQuality} and Fig.~\ref{fig:ablation_structure} show the quantitative and visual results of the ablation study of two key components--residual map mechanism and domain adaptation.
	Compared with RSGAN (RM) and RSGAN (nRM), the LPIPS is reduced by nearly $29\times10^{-2}$, the PSNR and SSIM increase of 2.65 and 0.008, respectively, which illustrate that introducing residual map mechanism can improve the performance of inter-rib details.
	The metrics of RSGAN (RM) are also better than those of U-Net (DRR), which shows that our proposed disentangled network has better ability of generation than that of U-Net.
	After introducing the domain adaptation structure, RSGAN (RMDA) achieves lower Weber contrast of $2.45\times10^{-2}$ and lower $\mathbf{MAE}_{rec}$ of 0.027 than those of RSGAN (RM), and has much better $\mathbf{MAE}_{cyc}$ of 0.041 than that of U-Net (Cycle) and \citet{LiTMI}.
	It may because the proposed domain adaptation method can reduce the impact of domain variance between DRR and CXR, and can avoid the changing of the inter-rib area in the procedure of domain transform.
	
	Table~\ref{tab:ablation} shows the quantified comparisons over the different extra adaptive loss functions. The first row lists the baseline results from the typical adversarial learning loss function. We can see that, adding the suppressed loss  $\mathcal{L}_{G_{\bigtriangledown}}$ can reduce the Weber contrast by $0.74\times10^{-2}$, illustrating that the intensity difference decreases between the inner region of rib and background. Despite this, $\mathcal{L}_{G_{\bigtriangledown}}$ results in the decrease of PSNR and the increase of LPIPS, indicating the loss of details during rib suppression.
	Further introducing the inter-rib constraint loss $\mathcal{L}_{inter}$, LPIPS, PSNR and SSIM achieve obvious improvement, suggesting that $\mathcal{L}_{inter}$ facilitates the preservation of the inter-rib details when suppressing ribs.
	Additionally, the complementary Laplacian-based loss $\mathcal{L}_{\bigtriangleup}$ can result in marginal improvement on the listed metrics and retains clearer details in the inner-rib area of the rib-suppressed result.
	
\begin{figure*}[!htbp]
		\centering
		\includegraphics[scale=.9]{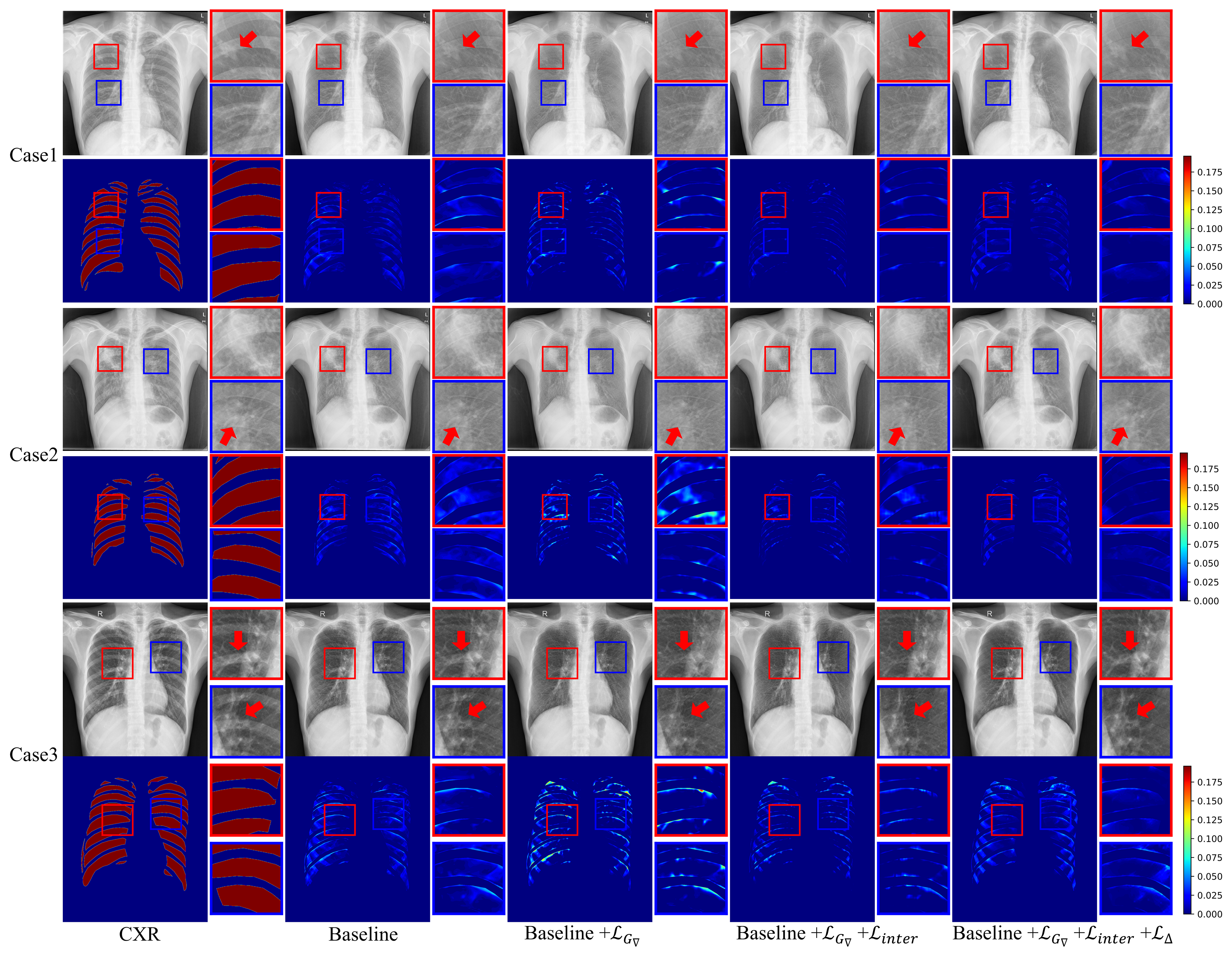}
		\caption{Rib-suppressed results for ablation study of extra adaptive loss functions. The first line of each case is the CXR images, and the second line is the colormap for the difference in the inter-rib area between the rib-suppressed image and the original CXR. The red region in the CXR colormap indicates the manually labeled mask of the inter-rib area. Rectangles in different colors refer to different areas in the image, and the areas are magnified at the right side of the image. Red arrows indicate some notable differences in the local area.}
		\label{fig:ablation}
	\end{figure*}

	As shown in Fig.~\ref{fig:ablation}, three CXRs are chosen from Shenzhen hospital dataset~\citep{MC&shenzhen} to visually compare the rib-suppressed performance by utilizing different loss functions.
	When adding with the $\mathcal{L}_{G_{\bigtriangledown}}$, we obtain the rib-suppressed result of CXR with fewer rib residues than that in the baseline, but with fewer details remaining.
	Comparing with $\mathcal{L}_{G_{\bigtriangledown}}$, $\mathcal{L}_{inter}$ helps keep the intensity in inter-rib area unchanged, and $\mathcal{L}_{\bigtriangleup}$ improves the preserving of the details in the inner-rib area.
	
	\begin{table*}[!t]
		\caption{\label{tab:SuppressionQuality}The quantitative results of CXR rib suppression of manual annotated image from Shenzhen hospital and Chest-xray14 dataset. The comparison methods involve state-of-art methods and ablation study with different key components in RSGAN (nRM indicates not using a residual map, RM indicates using a residual map, and RMDA indicates using both residual map and domain adaption). The best result is in bold and the second best one is underlined.}
		\centering
		\begin{tabular}{lcccccc}
			\hline
			Method & $C_w$ ($10^{-2}$)$\downarrow$ & LPIPS ($10^{-2}$)$\downarrow$ & PSNR$\uparrow$ & SSIM$\uparrow$ & $\mathbf{MAE}_{rec}\downarrow$ & $\mathbf{MAE}_{cyc}\downarrow$ \\
			\hline
			CXR & 4.04 & - & - & - & - & - \\
			U-Net (DRR) & 3.49 & 2.96 & 26.73 & 0.982 & - & - \\
			U-Net (Cycle) & 2.76 & 8.19 & 25.31 & 0.959 & - & 0.055 \\
			\citet{LiTMI} & \underline{2.36} & \underline{1.18} & \underline{34.17} & \underline{0.995} & - & 0.064 \\
			\hline
			RSGAN (nRM) & 3.53 & 31.5 & 25.09 & 0.978 & - & - \\
			RSGAN (RM) & 2.92 & 2.81 & 27.74 & 0.986 & 0.032 & - \\
			RSGAN (RMDA) & 2.45 & 1.51 & 30.15 & 0.992 & \underline{0.027} & \underline{0.041} \\
			RSGAN & \textbf{1.96} & \textbf{0.62} & \textbf{34.62} & \textbf{0.996} & \textbf{0.026} & \textbf{0.039} \\
			\hline
		\end{tabular}
	\end{table*}
	
	\begin{table*}[!htpb]
		\caption{\label{tab:ablation}Ablation study with different extra adaptive loss functions. The best result is in bold and the second best one is underlined.}
		\centering
		\begin{tabular}{cccccccc}
			\hline
			Baseline & $\mathcal{L}_{G_{\bigtriangledown}}$ & $\mathcal{L}_{inter}$ & $\mathcal{L}_{\bigtriangleup}$ & $C_w$ ($10^{-2}$)$\downarrow$ & LPIPS ($10^{-2}$)$\downarrow$ & PSNR$\uparrow$ & SSIM$\uparrow$ \\
			\hline
			\checkmark &  &  &  & 2.45 & 1.51 & 30.15 & 0.992 \\
			\checkmark & \checkmark &  &  & \textbf{1.71} & 2.08 & 28.54 & 0.989 \\
			\checkmark & \checkmark & \checkmark &  & 2.05 & \underline{0.83} & \underline{34.26} & \textbf{0.996} \\
			\checkmark & \checkmark & \checkmark & \checkmark & \underline{1.96} & \textbf{0.62} & \textbf{34.62} & \textbf{0.996} \\
			\hline
		\end{tabular}
	\end{table*}

    \subsection{Statistical tests}
\begin{figure*}[!htbp]
    	\centering
    	\includegraphics[scale=.9]{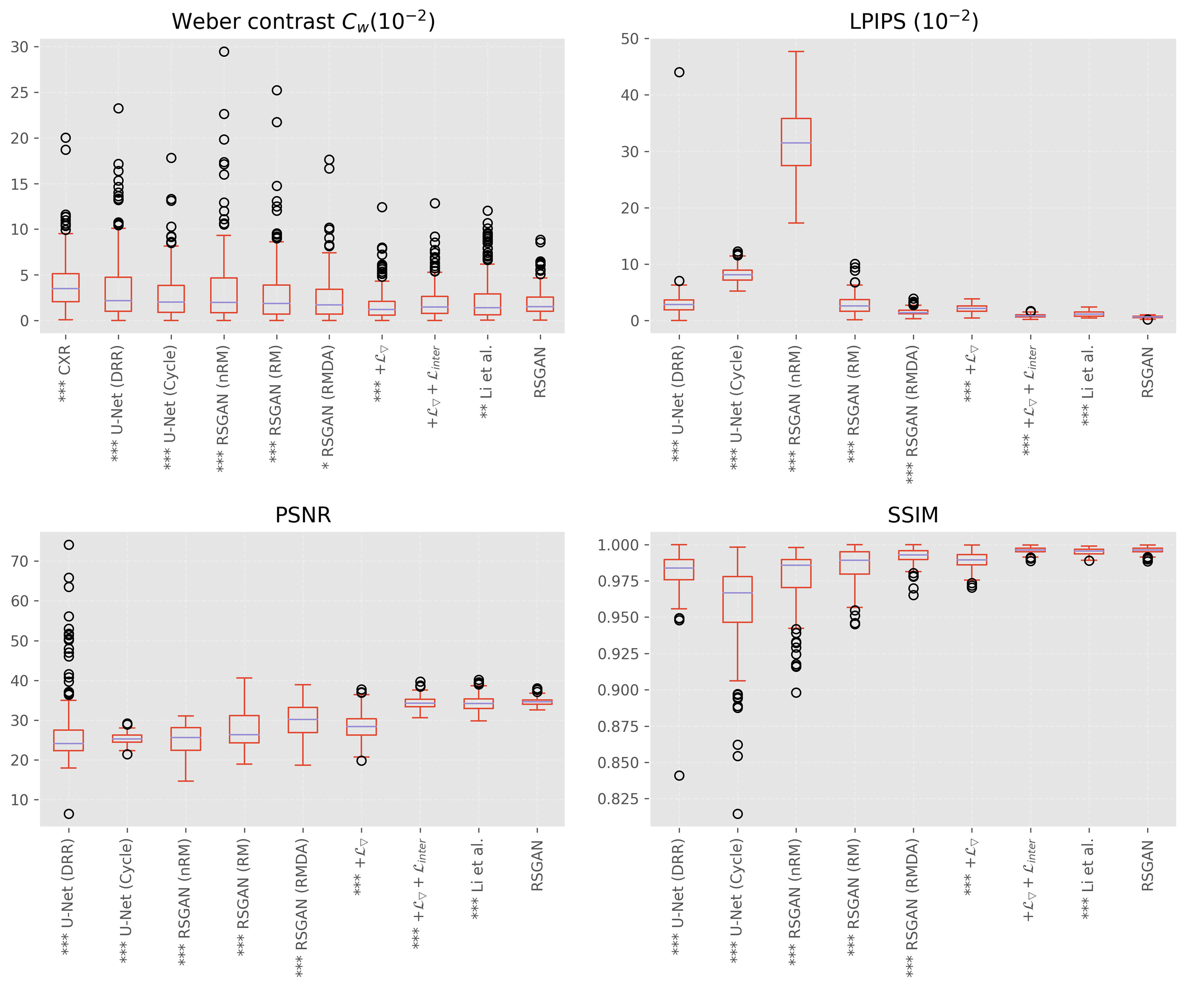}
    	\caption{Box plots of Weber contrast, LPIPS, PSNR, and SSIM for different methods. For statistically significant improvements between RSGAN and compared methods, methods marked with *, **, and *** indicate $0.01<p<0.05$, $0.001<p<0.01$, and $p<0.001$, respectively.}
    	\label{fig:statis}
    \end{figure*}

    Fig.~\ref{fig:statis} illustrates the box plots for Weber contrast, LPIPS, PSNR, and SSIM of RSGAN and compared methods. Paired T-tests show that the Weber contrast, LPIPS, PSNR and SSIM improvement given by RSGAN are statistically significant ($p<0.05$) against U-Net (DRR), U-Net (Cycle), \citet{LiTMI}, RSGAN (nRM), RSGAN (RM), RSGAN (RMDA) and "$+\mathcal{L}_{G_{\bigtriangledown}}$". Although RSGAN does not give statistically significant improvements on Weber contrast and SSIM against "$+\mathcal{L}_{G_{\bigtriangledown}}+\mathcal{L}_{inter}$" ($p=0.141$ and $p=0.900$), RSGAN achieves statistically significant improvements on LPIPS and PSNR ($p<0.05$).
    
	\begin{table*}[!t]
		\caption{\label{tab:chest14AUC}The area under curve (AUC) of predicting 14 lung diseases on Chest-xray14. The best result is in bold and the second best one is underlined.}
		\centering
		\begin{tabular}{lccccc}
			\hline
			\multirow{2}{*}{Method} & DenseNet-121 & DenseNet-121 & DenseNet-121 & DenseNet-121 & DenseNet-121  \\
			& ~\citep{ChestX-Ray8} & + \citet{LiTMI} & + \citet{LiTMI} (Mix) & + RSGAN & + RSGAN (Mix) \\
			\hline
			Atelectasis & 0.816 & 0.809 & \underline{0.817} & 0.815 & \textbf{0.827} \\
			Cardiomegaly & 0.901 & 0.902 & \textbf{0.906} & 0.900 & \textbf{0.906} \\
			Effusion & 0.877 & 0.881 & \underline{0.883} & \textbf{0.884} & 0.881 \\
			Infiltration & 0.705 & 0.698 & 0.701 & \underline{0.706} & \textbf{0.717} \\
			Mass & 0.838 & 0.832 & \textbf{0.860} & 0.839 & \underline{0.859} \\
			Nodule & 0.774 & 0.738 & \underline{0.786} & 0.749 & \textbf{0.795} \\
			Pneumonia & 0.753 & 0.759 & \textbf{0.770} & 0.759 & \underline{0.769} \\
			Pneumothorax & 0.868 & 0.867 & \textbf{0.884} & 0.878 & \textbf{0.884} \\
			Consolidation & 0.807 & 0.805 & \textbf{0.812} & \underline{0.810} & 0.809 \\
			Edema & 0.892 & 0.891 & \underline{0.896} & 0.894 & \textbf{0.900} \\
			Emphysema & 0.925 & 0.914 & 0.931 & \underline{0.934} & \textbf{0.935} \\
			Fibrosis & 0.831 & 0.828 & 0.844 & \underline{0.847} & \textbf{0.850} \\
			Pleural Thickening & \underline{0.783} & 0.770 & 0.778 & 0.776 & \textbf{0.784}\\
			Hernia & 0.886 & \textbf{0.954} & 0.940 & \underline{0.942} & \underline{0.942} \\
			\textbf{Average} & 0.833 & 0.832 & \underline{0.844} & 0.838 & \textbf{0.847} \\
			\hline
		\end{tabular}
	\end{table*}
	
	\begin{table*}[!t]
		\caption{\label{tab:TBacc}CXR image classification results on TBX11K test set. The best result is in bold and the second best one is underlined.}
		\centering
		\begin{tabular}{lcccccc}
			\hline
			Method & Accuracy & AUC (TB) & Sensitivity & Specificity & Ave. Prec. & Ave. Rec. \\
			\hline
			FRCNN~\citep{liu2020rethinking} & 86.49 & 94.83 & 75.58 & 94.35 & 84.72 & 84.85 \\
			FRCNN + \citet{LiTMI} & 86.95 & 95.21 & 78.67 & 93.52 & 84.94 & 85.75 \\
			FRCNN + \citet{LiTMI} (Mix) & \underline{87.83} & \underline{95.54} & \underline{81.30} & 93.11 & 85.77 & \underline{87.06} \\
			FRCNN + RSGAN & 87.28 & 95.39 & 75.73 & \textbf{95.86} & \underline{86.30} & 85.66 \\
			FRCNN + RSGAN (Mix) & \textbf{89.19} & \textbf{96.32} & \textbf{83.31} & \underline{95.25} & \textbf{88.04} & \textbf{88.65} \\
			\hline
		\end{tabular}
	\end{table*}
	
	\begin{table*}[!htpb]
		\caption{\label{tab:TBap}TB area detection results on TBX11K test set. CA TB denotes class-agnostic TB. The best result is in bold and the second best one is underlined.}
		\centering
		\begin{tabular}{lcccccc}
			\hline
			\multirow{2}{*}{Method} & \multicolumn{2}{c}{CA TB} & \multicolumn{2}{c}{Active TB} & \multicolumn{2}{c}{Latent TB} \cr\cline{2-7} & AP$_{50}$ & mAP & AP$_{50}$ & mAP & AP$_{50}$ & mAP \\
			\hline
			FRCNN~\citep{liu2020rethinking} & 55.76 & \textbf{25.65} & \underline{51.35} & \textbf{23.55} & 5.81 & 1.71 \\
			FRCNN + \citet{LiTMI} & 50.77 & 20.61 & 46.69 & 18.80 & \underline{6.99} & \underline{1.78} \\
			FRCNN + \citet{LiTMI} (Mix) & 55.67 & 22.66 & 51.28 & 21.21 & 3.86 & 1.08 \\
			FRCNN + RSGAN & \underline{55.77} & 23.95 & 50.56 & 22.07 &5.35 & 1.58 \\
			FRCNN + RSGAN (Mix) & \textbf{60.28} & \underline{24.59} & \textbf{54.14} & \underline{22.11} & \textbf{8.52} & \textbf{2.36} \\
			\hline
		\end{tabular}
	\end{table*}
	
	\subsection{Downstream applications}
	We employ two downstream applications, including lung disease classification and tuberculosis detection, to evaluate the quality of the imputed rib-suppressed CXR images.
	
	\subsubsection{Lung disease classification}
	To quantify the contribution of rib suppression in lung disease classification, we conduct experiments on Chest-xray14 dataset~\citep{ChestX-Ray8} and TBX11K~\citep{liu2020rethinking}.
	
	Referring to the experiments in \citep{ChestX-Ray8}, we utilize DenseNet-121~\citep{densenet} as classification network to predict 14 lung disease in Chest-xray14 dataset~\citep{ChestX-Ray8}, and conduct experiments on different input combinations.
	We propose five combinations including (1) DenseNet-121: only input with original CXR; (2) DenseNet-121 + Li: only input with rib-suppressed CXR generated by \citet{LiTMI}; (3) DenseNet-121 + Li (Mix): input with the concatenation of two original CXR channels and one rib-suppressed CXR channel generated by \citet{LiTMI}; (4) DenseNet-121 + RSGAN: only input with rib-suppressed CXR generated by our method; (5) DenseNet-121 + RSGAN (Mix): input with the concatenation of two original CXR channels and one rib-suppressed CXR channel generated by our method.
	We compare the five input combinations using the common metrics of Area Under Curve (AUC) on predicting 14 lung diseases on Chest-xray14 dataset~\citep{ChestX-Ray8}.
	The results are illustrated in Table~\ref{tab:chest14AUC}, and our method achieves state-of-the-art results on lung disease classification.
	Concatenating original CXR and rib-suppressed images generated by our method obtains a performance boost of 0.014 in average AUC than single original CXR.
	Only utilizing our rib-suppressed CXR achieves 1\% AUC higher in pneumothorax and 1.6\% AUC higher in fibrosis than original CXR because these lung diseases cause appearance changes in lung textures and rib suppression helps highlight the lung texture.
	
	Referring to the experiments in \citet{liu2020rethinking}, we utilize Faster R-CNN (FRCNN) as a classification network to predict CXR into three classes (healthy, unhealthy but non-tuberculosis, tuberculosis).
	Similarly, we propose five combination including (1) FRCNN; (2) FRCNN + \citet{LiTMI}; (3) FRCNN + \citet{LiTMI} (Mix); (4) FRCNN + RSGAN; and (5) FRCNN + RSGAN (Mix).
	Here we compare the five input combinations using the metrics of Accuracy, AUC, Sensitivity, Specificity, Average Precision, and Average Recall on predicting three classes. The comparisons are based on TBX11K dataset~\citep{liu2020rethinking} using the official splits.
	As illustrated in Table~\ref{tab:TBacc}, RSGAN achieves state-of-the-art results on CXR image classification with an increase of 2.7\% in accuracy and 1.49\% in AUC than those of input without rib-suppressed images.
	It is supported that rib suppression can help improve CXR image classification accuracy for clinical diagnosis.
	
	\subsubsection{Tuberculosis detection}
	In order to demonstrate the effectiveness of RSGAN for tuberculosis (TB) area detection, we utilize FRCNN as a detection network to predict TB area and classify each bounding box into category-agnostic tuberculosis (CA TB), active TB, and latent TB.
	Similar to classification tasks, we combine five types of input to train the network: (1) FRCNN; (2) FRCNN + \citet{LiTMI}; (3) FRCNN + \citet{LiTMI} (Mix); (4) FRCNN + RSGAN; and (5) FRCNN + RSGAN (Mix).
	
	For the evaluation of tuberculosis detection, we utilize two metrics of the average precision of the bounding box (AP).
	AP$_{50}$ refers to AP at the IoU (intersection-over-union) threshold of 0.5.
	mAP denotes the average AP with the IoU threshold from 0.5 to 0.95 with a step of 0.05.
	The higher the AP$_{50}$ and mAP are, the higher accuracy of tuberculosis detection the method gets.
	The comparisons are based on TBX11K dataset~\citep{liu2020rethinking} according to the official splits.
	The results are illustrated in Table~\ref{tab:TBap}.
	Combining the original CXR image with its rib-suppressed result can obtain the best detection results than the other types of input, and achieves approximately 5\% increase on AP$_{50}$ of CA TB than that of input only with original CXR image.
	It might be because that RSGAN could suppress the clavicles and ribs overlapped on the tuberculosis lesion, which is illustrated in the red rectangles of case 3 and case 4 in Fig.~\ref{fig:result}.
	
	
	\section{Discussion} \label{sec:discussion}
	In this study, we develop the RSGAN for suppressing ribs in the CXR images. Our proposed RSGAN is especially designed to handle with the challenges --- overlapped anatomies and unknown ground truth --- in CXR rib-suppression. 
	To obtain the rib structural features, we borrow the rib-suppression knowledge from DRR images. The structural-specific features of ribs are learned from our disentangled GAN framework. These features are then transferred to the CXR images for rib-suppression using the domain adaption technique. 
	Besides, to ensure the inter-rib intensities (i.e. those in the cavity regions) intact, we incorporate a residual map mechanism, which characterizes the intensity difference between a CXR and its corresponding rib-suppressed image, into domain adaption as a constraint for the contrast-specific features. Furthermore, extra adaptive loss functions are introduced to handle rib residue and preserve details in both inter-rib and inner-rib regions.
	
	We demonstrate the effectiveness of our proposed RSGAN by comparing with the state-of-the-art methods. Sufficient ablation studies are conducted to validate the improvement of each key component in RSGAN. 
	\paragraph{Residual Map---}
	ResNet~\citep{he2016deep} demonstrates that the residual representation can simplify the optimization for deep CNN.
	Many researches~\citep{nie2018medical,sun2020adversarial,de2021residual} prove that long-term residual connection benefits medical image synthesis. It is because that the residual maps have intensity distribution closed to zero, making CNN easier to be trained.
	
	Quantitative results listed in Table~\ref{tab:SuppressionQuality} and predictions shown in Fig.~\ref{fig:ablation_structure} prove that the residual map mechanism is efficient to improve the accuracy of predicting the rib-suppressed image.
	For the rib suppression task, the residual maps involve many pixels with the intensity of zero due to most areas except ribs on the image keep unchanged. On the other hand, the residual map is the combination of the projection for rib and tissue components, which is simpler than that of the rib-suppressed CXR.
	Thus the rib residual map is easier to be predicted than rib-suppressed CXR, which simplifies the task.
	As shown in Fig.~\ref{fig:result}, Fig.~\ref{fig:ablation_structure} and Table~\ref{tab:SuppressionQuality}, U-Net (DRR) and RSGAN (RM), the methods directly predict the residual map instead of the rib-suppressed image, achieve superior performance than that of RSGAN (nRM) from both statistical and visual results. Especially, the methods without using residual map are ineffective, resulting in blur boundaries and dropped details as shown in Fig.~\ref{fig:ablation_structure}.
	\paragraph{Domain Adaption---}
	Domain adaption is a key factor of suppressing the ribs in CXR images based on DRR images. In the proposed RSGAN, the structural features of DRR images are adaptively transferred to those of CXR images via the disentangled structure- and contrast-specific generators.
	CycleGAN~\citep{CycleGAN2017} is a widely used solution for domain adaption by constructing the mapping of the source-target domains. But the CycleGAN-based methods only learn the relationship between the CXR and DRR images and neglect the domain variance between original and rib-suppressed images.
	The disentanglement network provides a \textit{reconstruction consistency} for the process of image decomposition and reconstruction of a single image and a \textit{cycle consistency} for the domain transferring processes between CXR and DRR.
	Aided by these two consistency constraints, the methods equipped with a disentangled architecture achieve clearer details and less blur in the inter-rib area than U-Net (Cycle).
	It is demonstrated that disentangled domain adaptation can make up for the shortcoming of CycleGAN-based methods by implicit constraints.
	As shown in Table~\ref{tab:SuppressionQuality}, \citet{LiTMI} achieves higher $\mathbf{MAE}_{cyc}$ but better rib suppression performance than those of U-Net (Cycle). Worse $\mathbf{MAE}_{cyc}$ of \citet{LiTMI} is because that the LoG transformation is applied on both CXR and DRR images which can make bone component shaper and enlarge the consistency error.
	Different from U-Net (Cycle),
	\citet{LiTMI} avoids transferring the rib-suppressed image from DRR to CXR by generating a rib mask from domain-transferred image to solve the problem of the CycleGAN-based methods.
	But it is difficult to guarantee that the predicted binary rib masks are accurate at the boundary of the rib.
	As shown in Fig.~\ref{fig:result}, there is a plaque remained in the Case 2 of \citet{LiTMI}, which is caused by a wrong binary rib mask.
	
	\paragraph{Adaptive Loss Function---}
	An effective loss function is highly important for the convergence and performance of a neural network. The extra adaptive loss functions in our RSGAN are introduced to address the drawbacks in the method of \citet{LiTMI}.
	\citet{LiTMI} predicts the residual map from the bone component projection based on histogram matching and then substracts the residual map from the original CXR image constrained within the area of predicted binary mask.
	This method neglects the difference between the decomposed bone and the intensity residue and lacks a formulation to promote a consistent rib mask, which may result in sharp changes around rib edges.
	The loss function $\mathcal{L}_{G_{\bigtriangledown}}$ in RSGAN provides GAN-based learning to handle rib residue adaptively rather than binary segmentation. As shown in Table~\ref{tab:SuppressionQuality} and Table~\ref{tab:ablation}, baseline with $\mathcal{L}_{G_{\bigtriangledown}}$ achieves the lowest Weber Contrast than that of RSGAN and \citet{LiTMI}. However, LPIPS, PSNR, and SSIM perform worse when only introducing $\mathcal{L}_{G_{\bigtriangledown}}$. Weber Contrast is the metric to evaluate the average difference between the rib and its surrounding region. Because the rib-suppressed DRR images remain fewer details of the trachea than CXR, adversarial learning with $\mathcal{L}_{G_{\bigtriangledown}}$ blurs the details in the inter-rib area of CXR images, which leads to lower Weber contrast but worse LIPIS, PSNR, and SSIM.
	To solve this issue, $\mathcal{L}_{inter}$ and $\mathcal{L}_{\bigtriangleup}$ help keep the intensity in inter-rib area unchanged and improves the preserving of the details in the inner-rib area, respectively.
	
	Besides the performance of rib suppression, we also compare our proposed RSGAN with \citet{LiTMI} in downstream applications, including lung disease classification and tuberculosis detection.
	As shown in Table~\ref{tab:chest14AUC}, using only rib-suppressed image achieves a lower AUC than that of using only original CXR images. And combining CXR with its corresponding rib-suppressed image obtains better result than using either of them. It is demonstrated that the rib suppression can improve the reliability of pulmonary disease diagnosis by reducing overlapped anatomies and ambiguous structure details, but there might be some details missing in the rib suppressed image, which cause a drop of AUC in lung disease classification.
	Compared with the result of \citet{LiTMI}, combing the CXR with rib-suppressed image predicted by our proposed RSGAN obtains the best performance in both lung disease classification and tuberculosis detection tasks.
	
	\section{Conclusion} \label{sec:conclusion}
	In this paper, we propose a GAN-based disentanglement learning network to obtain the rib suppression result of a CXR image automatically. 
	The proposed approach aims to suppress ribs by predicting the residual map between the input CXR and its rib-suppressed image. 
	Specifically, we train the model on annotated DRRs for rib suppression and transfer structural priors derived from unpaired CT/DRR images into the CXR domain. 
	Furthermore, we propose three rib suppression loss functions based on prior knowledge to improve the performance of the generated residual map. 
	Experimental results based on multiple benchmarking CXR datasets demonstrate that the performances of automatic lung disease classification and TB area detection are boosted with the aid of rib-suppressed images produced by our approach.
	The limitation of our RSGAN is that minor pixel-level rib residues may appear in the predicted rib-suppressed CXR images caused by interpolation bias when resampling a low-resolution image to the high-resolution image. A future study will focus on the generation of a high-resolution residual map in order to achieve a more accurate rib-suppressed image.
	
	
	\bibliographystyle{model2-names.bst}\biboptions{authoryear}
	\bibliography{refs.bib}
	
	
	
	
\end{document}